\documentclass[12pt]{article}
\usepackage{amsfonts,bm}
\usepackage{graphicx}
\baselineskip
\offinterlineskip
\begin{document}

\begin{center}
{\bf Boolean Functions, Quantum Gates, Hamilton Operators, Spin Systems 
and Computer Algebra}
\end{center}

\begin{center}
{\bf Yorick Hardy$^\ast$ and Willi-Hans Steeb$^\dag$} \\[2ex]

$\ast$
Department of Mathematical Sciences, \\
University of South Africa, Johannesburg, South Africa, \\
e-mail: {\tt hardyy@unisa.ac.za}\\[2ex]

$\dag$
International School for Scientific Computing, \\
University of Johannesburg, Auckland Park 2006, South Africa, \\
e-mail: {\tt steebwilli@gmail.com}\\[2ex]
\end{center}

\strut\hfill

{\bf Abstract.} We describe the construction of quantum gates 
(unitary operators) from boolean functions and give
a number of applications. Both non-reversible and reversible boolean 
functions are considered. The construction of the Hamilton operator
for a quantum gate is also described with the Hamilton operator
expressed as spin system. Computer algebra implementations are provided.

\strut\hfill

\section{Introduction}

A boolean function $f$ on $n$ variables is a mapping
$\{ 0,1 \}^n$ into $\{ 0,1 \}$. Let $x_j \in \{ 0,1 \}$
for $j=1,\dots,n$. We set ${\bf x}=(x_1,x_2,\dots,x_n)$.
In the following $\cdot$ denotes the AND operation,
$+$ denotes the OR operation, $\oplus$ the XOR operation
and $\overline{\phantom{x}}$ is the NOT operation.
For $n=1$ we have the four boolean functions $f_1(x)=0$, $f_2(x)=1$, 
$f_3(x)=x$, $f_4(x)=\bar x$. The last two are of course reversible.
\newline

Let $X=\{ 0,1 \}$ and $x_j \in X$. A boolean function $\bf f$ 
with $n$ input variables, $x_1$, \dots, $x_n$ and $n$ output
variables, $y_1$, \dots, $y_n$ is a function ${\bf f}:X^n \to X^n$
obeying 
$$
{\bf f}(x_1,\dots,x_n)\mapsto(y_1,\dots,y_n).
$$
Here $(x_1,\dots,x_n) \in X^n$ is called 
the input vector and $(y_1,\dots,y_n) \in X^n$ is called the output
vector. An $n$-input and $n$-output boolean function 
${\bf f}$ is reversible if it maps each input 
vector to a unique output vector, i.e. the map is a bijection.
\newline

Quantum gates are described by unitary operators. In the finite
dimensional Hilbert space ${\mathbb C}^d$ we have $d \times d$
unitary matrices. We describe how $2^{n+1} \times 2^{n+1}$ unitary 
matrices can be associated with a non-reversible boolean function $f$ 
and how $2^{n} \times 2^{n}$ unitary matrices can be associated with 
reversible boolean functions $\bf f$. Furthermore the construction
of the associated Hamilton operator is derived as well as the
finding of the associated spin system.
\newline

Finally computer algebra implementations in SymbolicC++
for the two cases are provided.

\section{Reversible Boolean Function and Quantum Gates}

Let ${\bf f}(x_1,\dots,x_n)=(f_1(x_1,\dots,x_n),\dots,f_n(x_1,\dots,x_n))$
be a reversible boolean function where ${\bf f}:\{0,1\}^n\to\{0,1\}^n$ and
$f_j:\{0,1\}^n\to\{0,1\}$ for $j\in\{1,2,\ldots,n\}$. We consider the standard
basis in the Hilbert space ${\mathbb C}^2$
$$
\left\{\,\,
|0\rangle = \pmatrix { 1 \cr 0 },\,\,
|1\rangle = \pmatrix { 0 \cr 1 }\,\,\right\}.
$$
Then a corresponding $2^n \times 2^n$
permutation matrix $U_{\bf f}$ can be constructed from
$$
U_{\bf f}(|x_1\rangle \otimes |x_2\rangle \otimes \cdots \otimes |x_n\rangle) =
|f_1({\bf x})\rangle \otimes |f_2({\bf x})\rangle \otimes \cdots \otimes 
|f_n({\bf x})\rangle
$$
where ${\bf x}=(x_1,\dots,x_n)$.
Given the permutation matrix $U_{\bf f}$ the reversible
boolean function can be constructed as follows
$$
{\bf f}(x_1,\dots,x_n)=(y_1,\ldots,y_n)
\quad\Leftrightarrow\quad
(\langle y_1|\otimes\cdots\otimes\langle y_n|)
U_{\bf f}
(|x_1\rangle\otimes\cdots\otimes|x_n\rangle)=1.
$$
Let the function $b:\{0,1\}^n\to\{0,1,\ldots,2^n-1\}$ be defined by
$$
b(x_1,\dots,x_n):=\sum_{j=1}^n x_j 2^{n-j}.
$$
Then
$$
(U_{\bf f})_{b(x_1,\dots,x_n),b(y_1,\dots,y_n)}
=\left\{\matrix{1&{\bf f}(x_1,\dots,x_n)=(y_1,\ldots,y_n)\cr
0&\textrm{otherwise}}\right.
$$
i.e. the permutation matrix $U_{\bf f}$ has a 1 in row $b(y_1,\ldots,y_n)$
and column $b(x_1,\ldots,x_n)$ if and only if
${\bf f}(x_1,\dots,x_n)=(y_1,\ldots,y_n)$, otherwise
it has a 0 in that entry.

\section{Examples for Reversible Boolean Gates}

Example 1. Consider the reversible gate (Feynman gate)
$$
x_1 \to x_1, \qquad x_2 \to x_1 \oplus x_2.
$$
The inverse function is given by $(x_1,x_2) \to (x_1,x_1 \oplus x_2)$.
Let $|0\rangle$, $|1\rangle$ be the standard basis in
the Hilbert space ${\mathbb C}^2$. Thus we are looking
for the unitary matrix which implements $(x_1,x_2 \in \{ 0,1 \})$ 
$$
|x_1\rangle \otimes |x_2\rangle \mapsto 
|x_1\rangle \otimes |x_1 \oplus x_2\rangle.
$$
We have
$$
|0\rangle \otimes |0\rangle \mapsto |0\rangle \otimes |0\rangle, \quad
|0\rangle \otimes |1\rangle \mapsto |0\rangle \otimes |1\rangle
$$
$$
|1\rangle \otimes |0\rangle \mapsto |1\rangle \otimes |1\rangle, \quad
|1\rangle \otimes |1\rangle \mapsto |1\rangle \otimes |0\rangle. 
$$
This provides the $4 \times 4$ permutation matrix
$$
U = \pmatrix { 1 & 0 & 0 & 0 \cr 0 & 1 & 0 & 0 \cr
               0 & 0 & 0 & 1 \cr 0 & 0 & 1 & 0 }
\equiv \pmatrix { 1 & 0 \cr 0 & 1 } \oplus \pmatrix { 0 & 1 \cr 1 & 0 }
$$
which is the CNOT-gate and $\oplus$ denotes the direct sum. 
\newline

Example 2. Let $x_1,x_2 \in \{ 0,1 \}$ and $\oplus$ be the 
XOR operation. Then 
$$
(x_1,x_2) \mapsto (x_1 \oplus 1,x_1 \oplus x_2)
$$
is a 2-bit reversible gate since
$$
(0,0) \mapsto (1,0), \quad (0,1) \mapsto (1,1), \quad
(1,0) \mapsto (0,1), \quad (1,1) \mapsto (0,0).
$$
Let $|0\rangle$, $|1\rangle$ be the standard basis in ${\mathbb C}^2$.
To find the $4 \times 4$ permutation matrix $P$ such that
$$
P(|x_1\rangle \otimes |x_2\rangle) = 
|x_1 \oplus 1\rangle \otimes |x_1 \oplus x_2\rangle
$$
we calculate the Kronecker products of the vectors. This provides 
the four equations for $P$
$$
P \pmatrix { 1 \cr 0 \cr 0 \cr 0 } = \pmatrix { 0 \cr 0 \cr 1 \cr 0 }, \quad
P \pmatrix { 0 \cr 1 \cr 0 \cr 0 } = \pmatrix { 0 \cr 0 \cr 0 \cr 1 },
$$
$$
P \pmatrix { 0 \cr 0 \cr 1 \cr 0 } = \pmatrix { 0 \cr 1 \cr 0 \cr 0 }, \quad
P \pmatrix { 0 \cr 0 \cr 0 \cr 1 } = \pmatrix { 1 \cr 0 \cr 0 \cr 0 }.
$$
Consequently we obtain the $4 \times 4$ permutation matrix
$$
P = \pmatrix { 0 & 0 & 0 & 1 \cr 0 & 0 & 1 & 0 \cr
               1 & 0 & 0 & 0 \cr 0 & 1 & 0 & 0 }
$$
with the eigenvalues $+1$, $-1$, $+i$, $-i$.
\newline   

Example 3. Given the $4 \times 4$ permutation matrix
$$
U = \pmatrix { 0 & 1 & 0 & 0 \cr 0 & 0 & 0 & 1 \cr
               1 & 0 & 0 & 0 \cr 0 & 0 & 1 & 0 }
$$
with the eigenvalues $+1$, $-1$, $+i$, $-i$.
Then we obtain the corresponding boolean function as follows. 
Since the matrix has $4=2^{2}$ rows, the function 
${\bf f}:\{0,1\}^2\to\{0,1\}^2$ has two arguments. The first column 
(i.e. the column numbered 0) has a 1 in the third row
(the row numbered 2) for which $b^{-1}(0)=(0,0)$ and
$b^{-1}(2)=(1,0)$. Thus $(0,0)\to(1,0)$. From the second column
$(0,1)\to(0,0)$. The third column provides $(1,0)\to(1,1)$
and the fourth column $(1,1)\to(0,1)$. Thus we have the map
$$
(0,0)\mapsto(1,0),\qquad (0,1)\mapsto(0,0),\qquad
(1,0)\mapsto(1,1),\qquad (1,1)\mapsto(0,1).
$$
The right hand side provides the boolean expression
$$
{\bf f}(x_1,x_2) = 
(\overline{x_1}\cdot\overline{x_2} + x_1 \cdot \overline{x_2},
x_1\cdot\overline{x_2}+x_1\cdot x_2)
=(\overline{x_2},x_1).
$$
Example 4. Consider the reversible 3-input/3-output gate given by
\begin{eqnarray*}
x_1' &=& x_1 \oplus x_3 \\
x_2' &=& x_1 \oplus x_2 \\
x_3' &=& (x_1 \cdot x_2) \oplus (x_1 \cdot x_3) \oplus (x_2 \cdot x_3).
\end{eqnarray*}
The inverse is given by 
\begin{eqnarray*}
x_1 &=& x_1'\cdot x_2'\cdot \overline{x_3'} + \overline{x_1'}\cdot x_3' + \overline{x_2'}\cdot x_3' \\
x_2 &=& \overline{x_1'}\cdot x_2'\cdot \overline{x_3'} + x_1'\cdot x_3' + \overline{x_2'}\cdot x_3' \\
x_3 &=& x_1'\cdot \overline{x_2'}\cdot \overline{x_3'} + \overline{x_1'}\cdot x_3' + x_2'\cdot x_3'. 
\end{eqnarray*}
The permutation matrix takes the form
$$
\pmatrix{
1 & 0 & 0 & 0 & 0 & 0 & 0 & 0\cr
0 & 0 & 0 & 0 & 1 & 0 & 0 & 0\cr
0 & 0 & 1 & 0 & 0 & 0 & 0 & 0\cr
0 & 1 & 0 & 0 & 0 & 0 & 0 & 0\cr
0 & 0 & 0 & 0 & 0 & 0 & 0 & 1\cr
0 & 0 & 0 & 1 & 0 & 0 & 0 & 0\cr
0 & 0 & 0 & 0 & 0 & 1 & 0 & 0\cr
0 & 0 & 0 & 0 & 0 & 0 & 1 & 0}.
$$

\section{Non-reversible Boolean Functions and Quantum Gates}

Let $|0\rangle$, $|1\rangle$ be the standard basis in the 
Hilbert space ${\mathbb C}^2$. Then the quantum states 
$$
|{\bf x}\rangle \equiv |x_1x_2\dots x_n\rangle \equiv 
|x_1\rangle \otimes |x_2\rangle \otimes \cdots \otimes |x_n\rangle 
$$
form a basis in the Hilbert space ${\mathbb C}^{2^n}$,
where $x_1,\dots,x_n \in \{ 0,1 \}$. We apply the ordering
$$
|000\dots 00\rangle, \,\,\, |000\dots 01\rangle, \,\,\, 
|000\dots 10\rangle, \dots, \,\,\, |111\dots 11\rangle.
$$ 
A general state $|\psi\rangle$ in the Hilbert space ${\mathbb C}^{2^n}$ 
can be written as
$$
|\psi\rangle = \sum_{j_1,\dots,j_n=0}^1 c_{j_1\dots j_n}
|j_1\rangle \otimes \cdots \otimes |j_n\rangle. 
$$ 
It is well-known (Nielsen and Chuang \cite{1}, Hardy and Steeb \cite{2}, 
Steeb and Hardy \cite{3}, Gruska \cite{4}, Hirvensalo \cite{5}, 
Mermin \cite{6}) that for a given boolean function $f$, there is a quantum 
circuit of comparable efficiency which computes the unitary transformation 
$U_f$ which takes as input the state $|{\bf x}\rangle \otimes |y\rangle$
($y \in \{ 0,1 \}$) in the Hilbert space ${\mathbb C}^{2^{n+1}}$ and gives 
the output state $|{\bf x}\rangle \otimes |y \oplus f({\bf x})\rangle$,
where $\oplus$ is the XOR-operation. Thus we can construct 
a $2^{n+1} \times 2^{n+1}$ unitary matrix such that
$$
U_f(|{\bf x}\rangle \otimes |y\rangle) =
|{\bf x}\rangle \otimes |y \oplus f({\bf x})\rangle.
$$
Owing to the selection of the standard basis the unitary matrix
$U_f$ will be a permutation matrix. Vice versa given a
$2^{n+1} \times 2^{n+1}$ permutation matrix, since we have
a reversible boolean function, we can construct the boolean function
using the same techniques as in the previous section.
\newline

If we start with the Hadamard basis in ${\mathbb C}^2$
$$
|0\rangle = \frac1{\sqrt2} \pmatrix { 1 \cr 1 }, \quad
|1\rangle = \frac1{\sqrt2} \pmatrix { 1 \cr -1 } 
$$
instead of the standard basis we obtain
$$
U_{f,H} = 
\left(U_H \otimes\cdots \otimes U_H\right)U_f
\left(U_H \otimes\cdots \otimes U_H\right)
$$
where $U_f$ is the permutation matrix that implements $f$ in the standard 
basis and $U_H$ is the Walsh Hadamard transform
$$
U_H = \frac1{\sqrt2} \pmatrix { 1 & 1 \cr 1 & -1 } = U_H^{-1}.
$$ 
Of course, this is just a change of basis.

\section{Examples for Non-Reversible Gates}

Example 1. Given the boolean function $f(x_1,x_2)=x_1 \cdot \bar x_2$. 
Thus the map is $(x_1,x_2,y \in \{ 0,1 \})$  
$$
|x_1\rangle \otimes |x_2\rangle \otimes |y\rangle \mapsto
|x_1\rangle \otimes |x_2\rangle \otimes |y \oplus (x_1 \cdot \bar x_2)\rangle 
$$
with 
$$
|0\rangle \otimes |0\rangle \otimes |0\rangle \mapsto 
|0\rangle \otimes |0\rangle \otimes |0\rangle, \quad
|0\rangle \otimes |0\rangle \otimes |1\rangle \to 
|0\rangle \otimes |0\rangle \otimes |1\rangle
$$
$$
|0\rangle \otimes |1\rangle \otimes |0\rangle \mapsto 
|0\rangle \otimes |1\rangle \otimes |0\rangle, \quad
|0\rangle \otimes |1\rangle \otimes |1\rangle \mapsto 
|0\rangle \otimes |1\rangle \otimes |1\rangle
$$
$$
|1\rangle \otimes |0\rangle \otimes |0\rangle \mapsto 
|1\rangle \otimes |0\rangle \otimes |1\rangle, \quad
|1\rangle \otimes |0\rangle \otimes |1\rangle \mapsto 
|1\rangle \otimes |0\rangle \otimes |0\rangle
$$
$$
|1\rangle \otimes |1\rangle \otimes |0\rangle \mapsto 
|1\rangle \otimes |1\rangle \otimes |0\rangle, \quad
|1\rangle \otimes |1\rangle \otimes |1\rangle \mapsto 
|1\rangle \otimes |1\rangle \otimes |1\rangle. 
$$
This leads to the $8 \times 8$ permutation matrix
$$
U_f = I_4 \oplus \pmatrix { 0 & 1 \cr 1 & 0 } \oplus I_2
$$
where $\oplus$ denotes the direct sum and $I_n$ is
the $n \times n$ identity matrix.
\newline

Example 2. Consider the boolean function $f(x_1,x_2)=x_1 \oplus x_2$, 
where for the XOR operation $0 \oplus 0=0$, $0 \oplus 1=1$, 
$1 \oplus 0=1$, $1 \oplus 1=0$. Then we have
$$
|0\rangle \otimes |0\rangle \otimes |0\rangle \mapsto 
|0\rangle \otimes |0\rangle \otimes |0\rangle, \quad
|0\rangle \otimes |0\rangle \otimes |1\rangle \mapsto 
|0\rangle \otimes |0\rangle \otimes |1\rangle
$$
$$
|0\rangle \otimes |1\rangle \otimes |0\rangle \mapsto 
|0\rangle \otimes |1\rangle \otimes |1\rangle, \quad
|0\rangle \otimes |1\rangle \otimes |1\rangle \mapsto 
|0\rangle \otimes |1\rangle \otimes |0\rangle
$$
$$
|1\rangle \otimes |0\rangle \otimes |0\rangle \mapsto 
|1\rangle \otimes |0\rangle \otimes |1\rangle, \quad
|1\rangle \otimes |0\rangle \otimes |1\rangle \mapsto 
|1\rangle \otimes |0\rangle \otimes |0\rangle
$$
$$
|1\rangle \otimes |1\rangle \otimes |0\rangle \mapsto 
|1\rangle \otimes |1\rangle \otimes |0\rangle, \quad
|1\rangle \otimes |1\rangle \otimes |1\rangle \mapsto 
|1\rangle \otimes |1\rangle \otimes |1\rangle. 
$$
This provides the $8 \times 8$ permutation matrix
$$
U_f = I_2 \oplus \pmatrix { 0 & 1 \cr 1 & 0 } \oplus 
\pmatrix { 0 & 1 \cr 1 & 0 } \oplus I_2
$$
where $\oplus$ denotes the direct sum.
\newline

Example 3. Let $n=3$. Consider the majority gate
$$
f(x_1,x_2,x_3) = (x_1 \cdot x_2) + (x_1 \cdot x_3) + (x_2 \cdot x_3)
$$
which returns 1 if two or three arguments of $f$ are 1 and 0 otherwise.
Then we obtain the $16 \times 16$ permutation matrix
$$
U_f = I_6 \oplus \pmatrix { 0 & 1 \cr 1 & 0 } \oplus I_2
\oplus\left(I_3\otimes \pmatrix { 0 & 1 \cr 1 & 0 }\right).
$$

\section{Construction of Hamilton Operators}

For any unitary matrix $V$ one can find a skew-hermitian matrix 
$K$ with $V=e^K$. The skew-hermitian matrix can then be identified with 
a Hamilton operator $\hat H$ (hermitian matrix times $\hbar\omega$) 
via $K=-i\hat Ht/\hbar$ (Steeb and Hardy \cite{7}).
The eigenvalues of a unitary matrix are of the form $e^{i\phi}$
$(\phi \in {\mathbb R})$. Let $P$ be an $n \times n$ permutation matrix then
$P^T$ is a permutation matrix and $PP^T=I$, i.e. $P^T=P^{-1}$.
Any permutation matrix has an eigenvalue $+1$ with the corresponding
normalized eigenvector $\frac1{\sqrt{n}}(1 \, 1 \, \dots \, 1)^T$.
The construction of the skew-hermitian matrix $K$ may be done via
the spectral decomposition of $U$, i.e. we find the eigenvalues and
the normalized (pairwise orthonormal) eigenvectors of $U$. Notice
that $V=e^{K}=e^{K+2\pi ki I}$ for all $k\in\mathbb{Z}$, so $K$ is
not unique. More generally, if $J$ is any matrix which commutes
with $K$ and $e^J=I$, then $e^{K+J}=V$.
\newline

Example 1. For the CNOT gate given above we obtain a skew-hermitian matrix 
$$
K = \pmatrix { 0 & 0 \cr 0 & 0 } \oplus
\left( \frac{\pi i}{2} \pmatrix { 1 & -1 \cr -1 & 1 }\right) 
$$
where we utilized the spectral decomposition of $U$ to find $K$.
\newline

Example 2. For the $4 \times 4$ permutation matrix
$$
P = \pmatrix { 0 & 0 & 0 & 1 \cr 0 & 0 & 1 & 0 \cr
               1 & 0 & 0 & 0 \cr 0 & 1 & 0 & 0 }
$$
given above with the eigenvalues $+1$, $-1$, $+i$, $-i$ we obtain
a skew-hermitian matrix $K$ with $P=e^K$
$$
K = \frac{\pi}{4} 
\pmatrix { i & i & -1-i & 1-i \cr
           i & i & 1-i & -1-i \cr
           1-i & -1-i & i & i \cr
           -1-i & 1-i & i & i }
$$
or
$$
K = -i\frac{\pi}{4} 
\pmatrix { -1 & -1 & 1-i & 1+i \cr
           -1 & -1 & 1+i & 1-i \cr
           1+i & 1-i & -1 & -1 \cr
           1-i & 1+i & -1 & -1 }
$$
with $\omega t=\pi/4$.
\newline

The Pauli spin matrices 
$$
\sigma_1 = \pmatrix { 0 & 1 \cr 1 & 0 }, \quad
\sigma_2 = \pmatrix { 0 & -i \cr i & 0 }, \quad 
\sigma_3 = \pmatrix { 1 & 0 \cr 0 & -1 } 
$$
together with the $2 \times 2$ identity matrix $\sigma_0=I_2$
form an orthogonal basis in the vector space of $2 \times 2$ matrices.
The spin matrices $S_1$, $S_2$, $S_3$ are given by 
$S_1=\frac12\sigma_1$, $S_2=\frac12\sigma_2$, $S_3=\frac12\sigma_3$.
Thus each Hamilton operator in the Hilbert space ${\mathbb C}^{2^n}$
can be written as linear combinations of Kronecker products of
Pauli spin matrices $\sigma_0$, $\sigma_1$, $\sigma_2$, $\sigma_3$.
\newline

For example 2 given above we find the Hamilton operator 
$$
\hat H = \hbar\omega(-\sigma_0 \otimes \sigma_0 - \sigma_0 \otimes \sigma_1
+ \sigma_1 \otimes \sigma_0 + \sigma_2 \otimes \sigma_0 
+ \sigma_1 \otimes \sigma_1 - \sigma_2 \otimes \sigma_1).
$$   

\section{Computer Algebra Implementations}
 
Our computer algebra implementation uses the computer algebra system
SymbolicC++ \cite{8}. In addition to the methods described above we
also compute a symbolic expression for $\mathbf{f}(x_1,\ldots,x_n)$,
the sum of products form obtained from the constructed truth tables
(which we simplified using an implementation of resolution 
(Lloyd \cite{9})).\\

The program illustrates Example 1 of section 5. The function \verb|main|
first finds the permutation matrix implementing the example:
$f(x_1,x_2)=x_1\cdot\overline{x}_2$. Then the map (truth table) is printed.
Finally we recreate the map from the permutation matrix, which is the 
reversible map
$$
g(x_1,x_2,x_3)=(x_1,x_2,(x_1\cdot\overline{x}_2)\oplus x_3)
               =(x_1,x_2,x_1\cdot\overline{x}_2\cdot\overline{x_3}+\overline{x_1}\cdot x_3+x_2\cdot x_3).
$$
The program counts from 0, i.e. $x_0$. The output is
\small 
\begin{verbatim}
[1 0 0 0 0 0 0 0]
[0 0 0 0 0 1 0 0]
[0 0 1 0 0 0 0 0]
[0 0 0 1 0 0 0 0]
[0 0 0 0 1 0 0 0]
[0 1 0 0 0 0 0 0]
[0 0 0 0 0 0 1 0]
[0 0 0 0 0 0 0 1]
000 -> 000
001 -> 001
010 -> 010
011 -> 011
100 -> 101
101 -> 100
110 -> 110
111 -> 111
[                 x0                ]
[                 x1                ]
[x0*NOT[x1]*NOT[x2]+NOT[x0]*x2+x1*x2]
\end{verbatim}
\normalsize
The full program listing follows.
\small
\begin{verbatim}
#include <bitset>
#include <iostream>
#include <list>
#include <map>
#include <vector>
#include "symbolicc++.h"
using namespace std;

const int n=3;

// a class to provide ordering of bitsets
// so that they can be used in maps
template <const size_t n>
class cmpbst
{
 public:
  bool operator()(const bitset<n> &b1,const bitset<n> &b2)
  {
   size_t i;
   for(i=0;i<n;++i) if(b1[i] != b2[i]) return (b1[i] < b2[i]);
   return false;
  }
};

// for a given reversible boolean map, find the corresponding
// permutation matrix
template <const size_t n>
Symbolic permutation(const map<bitset<n>,bitset<n>,cmpbst<n> > &m)
{
 unsigned int N = (1 << n);
 Symbolic P = Symbolic("P",N,N)*0;
 typename map<bitset<n>, bitset<n> >::const_iterator i;
 for(i=m.begin();i!=m.end();++i)
   P(i->second.to_ulong(),i->first.to_ulong()) = 1;
 return P;
}

// simplifies a sum of products form using resolution the 
// products are represented by bitsets and the sum is the 
// list of bitsets
template <const size_t n>
list<pair<bitset<n>,bitset<n> > > simplify(const list<bitset<n> > &s)
{
 bool change = true;
 // a list which indicates whether bitsets were used in resolution
 // or need to be copied to the next round
 list<bool> copy;
 list<bool>::iterator ci1, ci2;
 // each bitset is stored with a mask which indicates which bits
 // may be used for resolution, once a bit is used it will be masked
 list<pair<bitset<n>,bitset<n> > > r, t1, t2, *tp1 = &t1, *tp2 = &t2, *tpp;
 typename list<bitset<n> >::const_iterator li;
 typename list<pair<bitset<n>, bitset<n> > >::const_iterator lpi1, lpi2;

 for(li=s.begin();li!=s.end();++li)
 {
  t1.push_back(make_pair(*li,bitset<n>()));
  // initially all bitsets propagate
  copy.push_back(true);
 }
 while(!tp1->empty())
 {
  // track whether resolution has been applied
  // if no change is recorded, we are done
  change = false;
  for(lpi1=tp1->begin(),ci1=copy.begin();lpi1!=tp1->end();++lpi1,++ci1)
  {
   // search for a second bitset which differs from this bitset
   // in exactly one place (taking into account the masks)
   for(lpi2=lpi1,ci2=ci1;lpi2!=tp1->end();++lpi2,++ci2)
   {
    // only compare if the masks are the same
    if(lpi1->second==lpi2->second)
    {
     // XOR finds the differing bits which are then masked
     bitset<n> diff = ((lpi1->first ^ lpi2->first) & ~lpi1->second);
     // only one bit differs so apply resolution
     if(diff.count()==1)
     {
      // mask the bit which has been used
      tp2->push_back(make_pair(lpi1->first,lpi1->second | diff));
      change = true;
      // these bitsets have been used in resolution, don't copy them
      *ci1 = *ci2 = false;
     }
    }
   }
   if(*ci1) r.push_back(*lpi1);
  }
  // reset the variables for the next application of resolution
  tpp = tp1; tp1 = tp2; tp2 = tpp; tp2->clear();
  copy.clear(); copy.resize(tp1->size(),true);
 }
 r.unique();
 return r;
}

// find a symbolic expression for a given boolean map
template <const size_t n>
Symbolic expression(const map<bitset<n>, bitset<n>, cmpbst<n> > &m)
{
 size_t j, k;
 Symbolic S("S",n), NOT("NOT"), x("x",n);
 vector<list<bitset<n> > > terms(n);
 vector<list<pair<bitset<n>, bitset<n> > > > simplified(n);
 typename map<bitset<n>,bitset<n> >::const_iterator i;
 typename list<pair<bitset<n>,bitset<n> > >::iterator li;
 // for each y_j, record all values of x_1,...,x_n such that y_j = 1
 for(i=m.begin();i!=m.end();++i)
  for(j=0;j<n;++j) if(i->second[j]) terms[j].push_back(i->first);
 // construct each symbolic expression for y_j
 for(j=0;j<n;++j)
 {
  S(j) = 0;
  // find a smaller set of terms
  simplified[j] = simplify(terms[j]);
  for(li=simplified[j].begin();li!=simplified[j].end();++li)
  {
   Symbolic P = 1;
   for(k=0;k<n;++k)
    if(!li->second[k])
    {
    // this is the usual construction of a product for the
    // sum of products form generated from a truth table 
    if(li->first[k]) P *= x(k); else P *= NOT[x(k)];
    }
   S(j) += P;
  }
 }
 return S;
}

// determine the reversible boolean map from a permutation matrix
template <const size_t n>
map<bitset<n>,bitset<n>,cmpbst<n> > booleanmap(const Symbolic &permutation)
{
 size_t i, j;
 map<bitset<n>, bitset<n>, cmpbst<n> > m;
 for(i=0;i<(1<<n);++i)
  for(j=0;j<(1<<n);++j)
   if(permutation(i,j)!=0) m[bitset<n>(j)] = bitset<n>(i);
 return m;
}

// reverse the contents of a bitset
template <const int n>
bitset<n> reverse(const bitset<n> &b)
{
 size_t i;
 bitset<n> r;
 for(i=0;i<n;++i) r[n-i-1] = b[i];
 return r;
}

int main(void)
{
 int i1, i2, i3;
 bitset<3> a, b;
 map<bitset<3>, bitset<3>, cmpbst<3> > f, g;
 map<bitset<3>, bitset<3>, cmpbst<3> >::const_iterator i;
 Symbolic P;
 for(i1=0;i1<2;++i1)
  for(i2=0;i2<2;++i2)
   for(i3=0;i3<2;++i3)
   {
   a[0] = b[0] = i1; a[1] = b[1] = i2; a[2] = i3;
   b[2] = a[2]^(a[0] & (!a[1]));
   f[a] = b;
   }
 P = permutation(f);
 cout << P << endl;
 g = booleanmap<3>(P);
 for(i=g.begin();i!=g.end();++i)
  cout << reverse<3>(i->first) << " -> " << reverse<3>(i->second) << endl;
 cout << expression(f) << endl;
 return 0;
}
\end{verbatim}
\normalsize

\section{Conclusion}

We have described the algorithms for finding the permutation matrices
which implement boolean functions as a quantum gate and vice versa.
The construction of the Hamilton operator and the corresponding
spin system from the quantum gate is
also described. Computer algebra implementations were demonstrated.

\strut\hfill

{\bf Acknowledgment}
\newline

The authors are supported by the National Research Foundation (NRF),
South Africa. This work is based upon research supported by the National
Research Foundation. Any opinion, findings and conclusions or recommendations
expressed in this material are those of the author(s) and therefore the
NRF do not accept any liability in regard thereto.

\strut\hfill

\end{document}